\documentclass[aps,pre,twocolumn,showpacs,preprintnumbers,a4paper]{revtex4} 

\usepackage{graphics}
\usepackage{epsf}
\usepackage{epsfig}

\begin{document} 

\flushbottom
\def\bottomfraction{0.5}

\title{Scaling behavior of the order parameter and its conjugated
field in an absorbing phase transition around the upper critical
dimension}

\author{S. L\"ubeck}
\email{sven@thp.uni.duisburg.de}
\affiliation{
Theoretische Tieftemperaturphysik,
Gerhard-Mercator-Universit\"at, 
Lotharstr. 1, 
47048 Duisburg, Germany}

%\date\today
\date{Dezember 12, 2001}

\begin{abstract}
We analyse numerically the critical behavior of an absorbing
phase transition in a
conserved lattice gas in an external field.
The external field is realized as a spontaneous creation
of active particles which drives the system away from
criticality.
Nevertheless, the order parameter obeys certain scaling laws
for sufficiently small external fields.
These scaling laws are investigated and the corresponding
exponents are determined in various dimensions ($D=2,3,4,5$).
At the so-called upper critical dimension $D_{\text c}=4$
one has to modify the usual scaling laws by logarithmic 
corrections.
\end{abstract}



\preprint{accepted for publication in {\it Physical Review E} 2002}

\maketitle

\section{Introduction}

Recently Rossi {\it et at.}~introduced a conserved
lattice gas (CLG) with a stochastic short range 
interaction that exhibits a continuous phase transition
from an active state to an absorbing non-active state
at a critical value of the particle density~\cite{ROSSI_1}.
The CLG model is expected to belong to a new 
universality class of absorbing phase transitions 
characterized by a conserved field.
Similar to the well known universality
hypothesis of directed percolation~\cite{JANSSEN_1,GRASSBERGER_2}
the authors conjectured that "all stochastic models with an 
infinite number of absorbing states in which the order
parameter evolution is coupled to a nondiffusive
conserved field define a unique universality 
class"~\cite{ROSSI_1}.
In order to check this hypothesis the scaling behavior,
i.e.~the values of the corresponding critical exponents,
has to be determined.
The order parameter exponent as well as the exponent of the
order parameter fluctuations were determined in~\cite{LUEB_19}.
Furthermore a modified CLG model with random particle hopping
was introduced in~\cite{LUEB_20} which mimics the mean-field 
scaling behavior of the system.

In this work we investigated for the first time 
the CLG model in an external field.
The external field is conjugated to the order parameter,
i.e.~it is realized as a spontaneous creation of active
particles.
Of course a spontaneous creation of active particles
destroys the absorbing state and thus the absorbing
phase transition at all.
But analogous to ferromagnetic phase transitions the 
order parameter can be described as a generalized homogeneous
function of the control parameter and of the conjugated field
in the critical regime. 
This scaling behavior is examined
below ($D=2,3$), above ($D=5$), and
at the so-called upper critical dimension ($D=4$).

\section{Model and scaling behavior}
\label{sec:model}

We consider the CLG model on $D$-dimensional cubic
lattices of linear size~$L$.
Initially one distributes randomly $N=\rho L$ 
particles on the system where $\rho$ denotes the
particle density.
In order to mimic a repulsive interaction
a given particle is considered as {\it active} if at least one of its 
$2D$ neighboring sites on the cubic lattice is occupied
by another particle.
If all neighboring sites are empty the particle
remains {\it inactive}.
Active particles are moved in the next update step to one 
of their empty nearest neighbor sites, selected at random.
In the steady state, the system is characterized by the
density of active sites $\rho_{\text a}$.
The density $\rho_{\text a}$ is the order parameter of 
the absorbing phase transition, i.e., it vanishes 
if the control parameter $\rho$ is lower
than the critical value $\rho_{\text c}$.
In the thermodynamic limit the order parameter 
scales for zero field and for $\rho > \rho_{\text c}$ as
\begin{equation}
\rho_{\text{a}}(\delta\rho, h=0) \; \sim \; 
\delta \rho^{\beta}
\label{eq:rho_order_par}
\end{equation}
with $\delta \rho =\rho/\rho_{\text{c}}-1$.

In this work we consider the CLG model in an 
external field~$h$ conjugated to the order parameter,
i.e., the field spontaneously creates active particles.
Clearly the particular implementation of the spontaneous creation of
active particles has to obey the particle
conservation of the CLG model which is believed to be a
relevant parameter, determining the universality class.
Instead of a creation of additional particles
we realize the external field by particle
movements which do not change the total number of particles. 
Therefore we choose randomly $L^D h$ particles on the lattice.
Each of these particles is then moved to one of its
empty next neighbors.
In this way inactive particles may be activated and the 
number of active sites is increased.

In our simulations we start with randomly distributed
particles.
The system is updated according to the above rules in the
following way.
One lists all active sites and updates these
sites in a randomly chosen sequence.
Then one moves $L^D h$ randomly
chosen inactive particles to one of its empty
next neighbors.
In the case that these movements create
active particles, these particles are added to the
list of active particles and will be updated in 
the following update step.
Thus one update step contains both, the update of
active sites and the additional movements of inactive
particles which mimics the external field.

\begin{figure}[t]
  \includegraphics[width=8.0cm,angle=0]{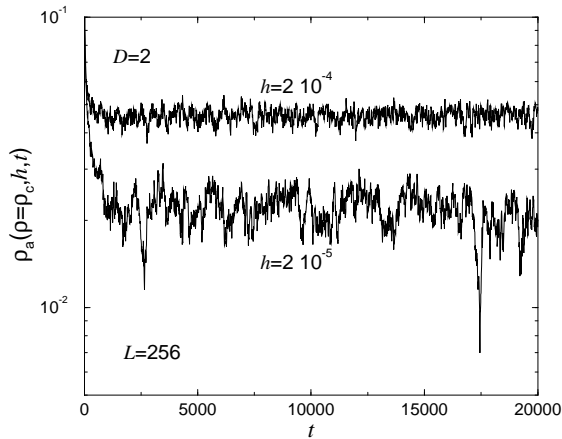}
  \caption{%%
    The density of active sites $\rho_{\text a}$ as a function
    of time (number of update steps) for two different values of 
    the external field~$h$.
    After a certain relaxation time the density of active sites
    fluctuates around a well defined value.
    For too small fields finite-size effects occur which results
    in strong fluctuation events (see lower curve).
   }
  \label{fig:rho_a_field_equil_01} 
\end{figure}

In Fig.\,\ref{fig:rho_a_field_equil_01}
we plot the density of active sites vs.~time (number of
update steps) at $\rho_{\text c}$ for two different values 
of the driving field.
After a certain relaxation time the system reaches a steady
state where the density of active sites fluctuates around
the average value $\langle \rho_{\text a}(\delta\rho,h,t)\rangle$
which is interpreted as the order parameter $\rho_{\text a}(\delta\rho,h)$.
Decreasing the external field below a certain value
finite-size effects occur which results in strong 
fluctuations.
Two of these strong fluctuations events can be seen in
the lower curve of Fig.\,\ref{fig:rho_a_field_equil_01}.
The origin of these effects is that the number of
particle movements ($L^D h$) create too few active
particles and the system tends to the absorbing state.
As a results the measured order parameter is shifted to
lower values.
To avoid these finite-size effects we increase the system size
before these fluctuation events occur.

The spontaneous creation of active particles destroys
the absorbing state and thereby the phase transition
itself.
Although the external field drives the system away from criticality
the order parameter obeys certain scaling laws
for sufficiently small fields.
At the critical density the order parameter is expected
to scale with the field as (see for instance~\cite{HINRICHSEN_1})
\begin{equation}
\rho_{\text{a}}(\delta\rho=0, h) \; \sim \; 
h^{\beta/\sigma}.
\label{eq:rho_field}
\end{equation}
As usual in critical phenomena the order parameter is 
assumed to be a generalized homogeneous function of the
control parameter $\delta\rho$ and the applied field~$h$
\begin{equation}
\rho_{\text{a}}(\delta\rho, h) \; = \; 
\lambda\, \, {\tilde r}
(\delta \rho \; \lambda^{-1/\beta}, h \; \lambda^{-\sigma/\beta})
\label{eq:scal_ansatz}
\end{equation}
with the scaling function ${\tilde r}$.
Choosing $\delta \rho \, \lambda^{-1/\beta}=1$ at zero field
one recovers Eq.\,(\ref{eq:rho_order_par}) whereas
$h \, \lambda^{-\sigma/\beta}=1$ leads to Eq.\,(\ref{eq:rho_field}) 
at the critical density, respectively.

Furthermore we consider the fluctuations of the order
parameter
\begin{equation}
\Delta \rho_{\text a}(\delta\rho, h) 
\; = \; L^D \, \left [ \langle \rho_{\text a}(\delta\rho, h,t)^2 \rangle
\, - \, \langle \rho_{\text a}(\delta\rho,h,t)\rangle^2 \right ].
\label{eq:def_fluc}
\end{equation}
For zero field the fluctuations are known to diverge approaching
the critical point~\cite{LUEB_19}
\begin{equation}
\Delta \rho_{\text a}(\delta\rho, h=0) 
\; \sim \; \delta\rho^{-\gamma^{\prime}}.
\label{eq:fluc_crit}
\end{equation}
The fluctuation exponent $\gamma^{\prime}$
fulfill the scaling relation~\cite{JENSEN_3}
\begin{equation}
\gamma^{\prime} \; = \; \nu_{\scriptscriptstyle \perp} D \, - \, 
2 \, \beta,
\label{eq:scal_rel_gamma_prime}
\end{equation}
where the exponent $\nu_{\scriptscriptstyle \perp}$
describes how the spatial correlation length diverges 
at the transition point.
In the critical regime we assume that the fluctuations
obey the scaling ansatz
\begin{equation}
\Delta \rho_{\text a}(\delta\rho, h) 
\; = \; 
\lambda^{\gamma^{\prime}}\, \, {\tilde d}
(\delta \rho \; \lambda, h \, \lambda^{\sigma})
\label{eq:fluc_scal_ansatz}
\end{equation}
Setting $\delta \rho \, \lambda=1$ one recovers 
Eq.\,(\ref{eq:fluc_crit}) for $h=0$.

Analogous to equilibrium phase transitions the susceptibility
is defined as the derivative of the order parameter with respect to the
conjugated field
\begin{eqnarray}
\label{eq:def_suscept}
\chi(\delta\rho,h) &  =  & \frac{\partial\hphantom{h}}{\partial h} \,
\rho_{\text a}(\delta\rho, h) \nonumber \\
& = &
\lambda^{1-\sigma/\beta}\, \, {\tilde c}
(\delta \rho \; \lambda^{-1/\beta}, h \; \lambda^{-\sigma/\beta}).
%\label{eq:def_suscept}
\end{eqnarray}
Approaching the transition point the susceptibility diverges for
zero field as
\begin{equation}
\chi(\delta\rho,h=0) \;  \sim  \; \delta\rho^{-\gamma}.
\label{eq:suscept_gamma}
\end{equation}
This result can be recovered from Eq.\,(\ref{eq:def_suscept}) by setting
$\delta\rho\lambda^{1/\beta}=1$ for $h=0$ and one gets the 
scaling relation
\begin{equation}
\gamma \; = \; \sigma \, - \,  \beta
\label{eq:widom}
\end{equation}
which corresponds to the well known Widom equation  
of equilibrium 
phase transitions.
Using this scaling relation
we calculate in the following the value of the 
susceptibility exponent~$\gamma$ from the obtained values
of $\beta$ and $\sigma$.
Notice that in contrast to the scaling behavior of 
equilibrium phase transitions
the non-equilibrium absorbing phase transition 
is characterized by $\gamma \neq \gamma^{\prime}$.

\section{Below the critical dimension}
\label{sec:d2_d3}

\subsection{D=2}
\label{subsec:d2}

\begin{figure}[t]
  \includegraphics[width=8.0cm,angle=0]{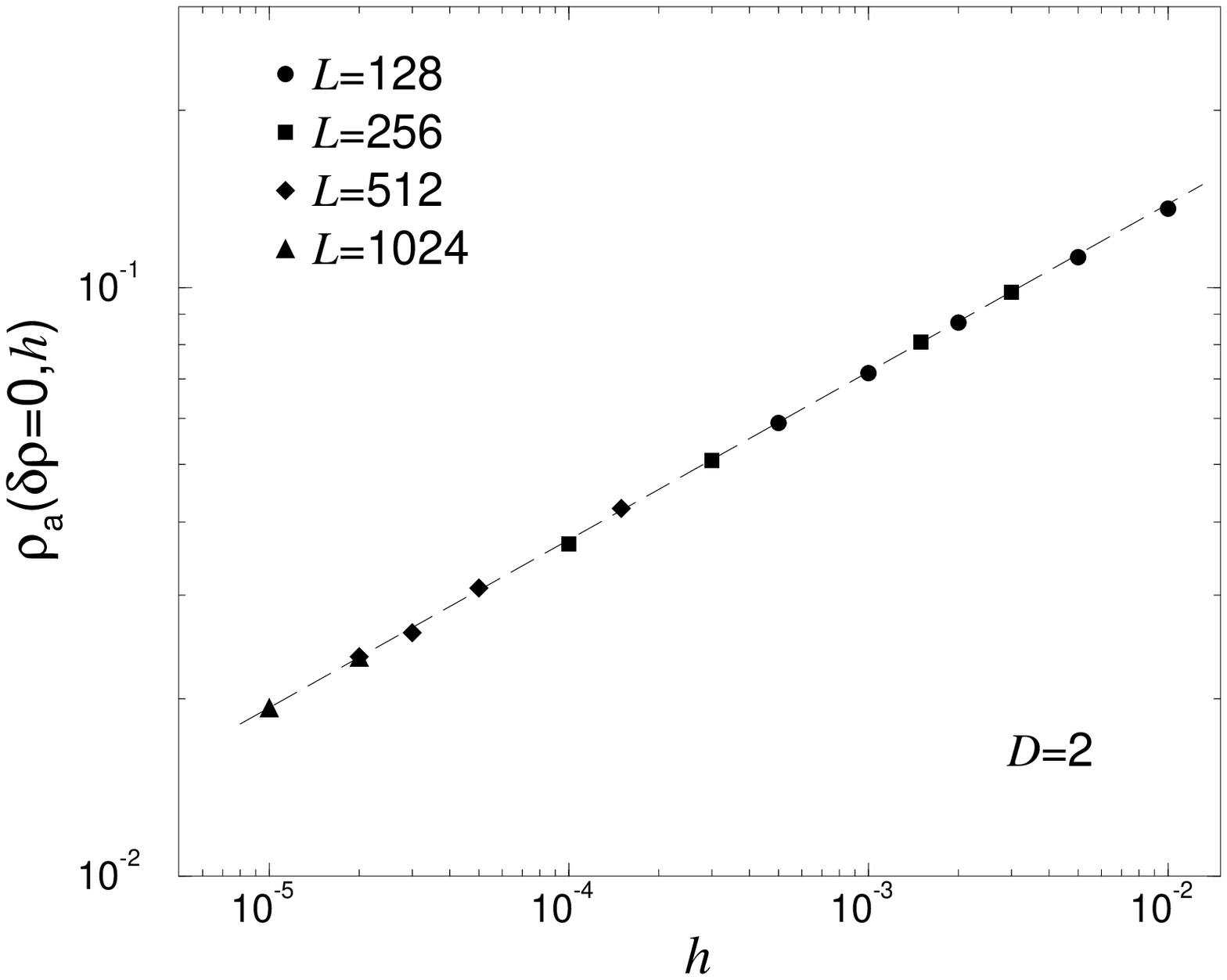}
  \caption{%%
    The order parameter~$\rho_{\text a}$ as a function of the
    field~$h$ at the critical density $\rho=\rho_{\text c}$ for $D=2$.
    The dashed line corresponds to a power-law behavior according
    to Eq.\,(\protect\ref{eq:rho_field}) with an exponent $0.286\pm0.001$.
   }
  \label{fig:rho_a_field_2d_01} 
\end{figure}

At the beginning of our analysis we consider the
two-dimensional CLG model at the critical density $\rho_{\text c}$
for various values of the driving field~$h$
where the value of the critical density is obtained
from~\cite{LUEB_19}. 
The field dependence of the order parameter 
$\rho_{\text a}(\delta\rho=0,h)$ 
is shown in a log-log plot Fig.\,\ref{fig:rho_a_field_2d_01}.
Approaching the critical point ($h\to 0$) 
the order parameter vanishes 
algebraically in agreement with Eq.\,(\ref{eq:rho_field}).
A regression analysis of the data yields the estimation
$\beta/\sigma=0.286\pm0.001$.
Using the value of the order parameter exponent
$\beta=0.637\pm0.009$~\cite{LUEB_19} we get $\sigma=2.227\pm0.032$.

\begin{figure}[b]
  \includegraphics[width=8.0cm,angle=0]{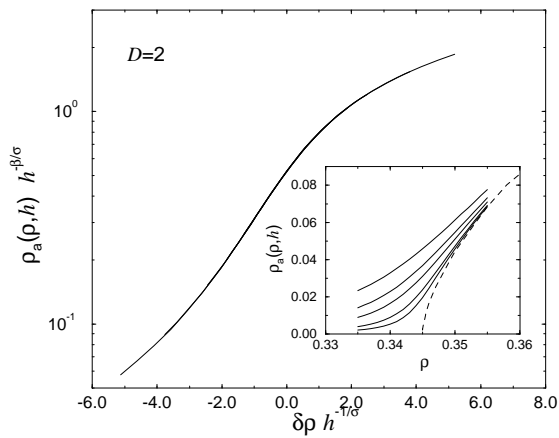}
  \caption{%%
    The scaling plot of the order parameter~$\rho_{\text a}$ for the
    two-dimensional model.
    The data are rescaled according to 
    Eq.\,(\protect\ref{eq:scal_plot_order}) with $\beta=0.637$ and
    $\sigma=2.227$.
    The inset displays the unscaled data, i.e., 
    the order parameter~$\rho_{\text a}$ is plotted as a 
    function of the density $\rho$ for different values 
    of the field~$h$.
    The order parameter is driven away from
    the critical point $(\rho_{\text c},0)$ with increasing field.
    The dashed line corresponds to the zero-field behavior.
   }
  \label{fig:rho_a_field_scal_2d_01} 
\end{figure}

In the following we analyse the order parameter as a function
of the control parameter $\delta\rho$ for different fields
from $h=10^{-5}$ up to $2\,10^{-4}$.
The applied field results in 
a rounding of the zero-field curve, i.e., the order parameter
increases smoothly with the control parameter for $h>0$
(see inset of Fig.\,\ref{fig:rho_a_field_scal_2d_01}).
According to the scaling ansatz of the order parameter
[Eq.\,(\ref{eq:scal_ansatz})] we choose $h \, \lambda^{-\sigma/\beta}=1$
and get the scaling form
\begin{equation}
\rho_{\text{a}}(\delta\rho, h) \; = \; 
h^{\beta/\sigma}\, \, {\tilde r}
(\delta \rho \; h^{-1/\sigma}, 1).
\label{eq:scal_plot_order}
\end{equation}
Thus plotting $\rho_{\text a} \, h^{-\beta/\sigma}$ as
a function of $\delta \rho \, h^{-1/\sigma}$ the curves
for different values of the driving field have to collapse
onto the scaling function ${\tilde r}$.
Using the above determined values of $\beta$ and $\sigma$ 
one gets an excellent data collapse which is 
shown in Fig.\,\ref{fig:rho_a_field_scal_2d_01}.
%The high quality of the obtained data collapse confirms
%the accuracy of the determination of $\rho_{\text c}$
%as well as of the critical exponents.

\begin{figure}[t]
  \includegraphics[width=8.0cm,angle=0]{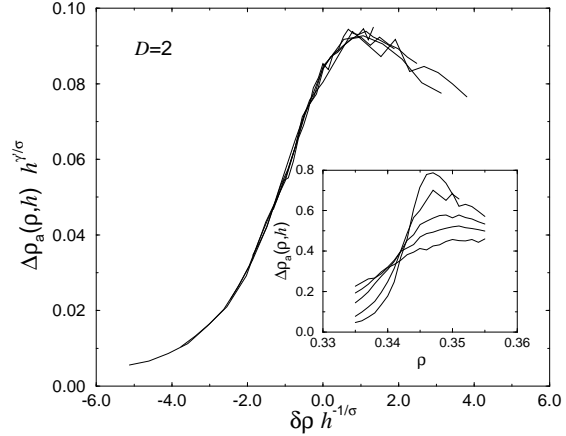}
  \caption{%%
    The scaling plot of the order parameter fluctuations~$\Delta\rho_{\text a}$ 
    for the two-dimensional model.
    The data are rescaled according to 
    Eq.\,(\protect\ref{eq:scal_plot_fluc}) with $\sigma=2.227$ and
    $\gamma^{\prime}/\sigma=0.18$.
    The inset displays the unscaled data, i.e., 
    the fluctuations are plotted as a 
    function of the density $\rho$ for different values 
    of the field~$h$.   
    Approaching the transition point ($h\to 0$ at $\rho=\rho_{\text c}$) 
    the peak of the fluctuations diverges.
   }
  \label{fig:fluc_a_scal_2d_01} 
\end{figure}

Next we consider the scaling behavior of the order
parameter fluctuations $\Delta\rho_{\text a}$.
%As usually the fluctuation data exhibits a stronger scattering 
%than the data of the order parameter.
The fluctuation data are shown for different values of the
external field in the inset of Fig.\,\ref{fig:fluc_a_scal_2d_01}.
For finite fields the fluctuations display a peak.
Approaching the transition point ($h\to 0$) this peak
becomes a divergence signalling the critical point.
In order to analyze the scaling behavior of the fluctuations 
we use Eq.\,(\ref{eq:fluc_scal_ansatz}) and set $h \, \lambda^{\sigma}=1$ 
which yields
\begin{equation}
\Delta \rho_{\text a}(\delta\rho, h) 
\; = \; 
h^{-\gamma^{\prime}/\sigma}\, \, {\tilde d}
(\delta\rho \; h^{-1/\sigma},1).
\label{eq:scal_plot_fluc}
\end{equation}
Since $\rho_{\text c}$ and $\sigma$ are already determined
one can estimate the exponent $\gamma^{\prime}$ by varying
this exponent until one observes a data collapse of the
different fluctuation curves.
The best result is obtained  for $\gamma^{\prime}/\sigma=0.18\pm0.02$,
leading to $\gamma^{\prime}=0.402\pm0.045$,
and the corresponding plot is shown in Fig.\ref{fig:fluc_a_scal_2d_01}.
The value agrees with $\gamma^{\prime}=0.384\pm0.023$~\cite{LUEB_19}
obtained from a regression analysis according to 
Eq.\,(\ref{eq:fluc_crit}).

Additionally the fluctuation exponent~$\gamma^{\prime}$ 
can be estimated via the scaling relation 
Eq.\,(\ref{eq:scal_rel_gamma_prime}).
Using the estimation $\nu_{\scriptscriptstyle \perp}=0.78\pm0.08$
we get $\gamma^{\prime}=0.286\pm 0.161$ which again agrees with our
result.

\subsection{D=3}
\label{subsec:d3}

\begin{figure}[t]
  \includegraphics[width=8.0cm,angle=0]{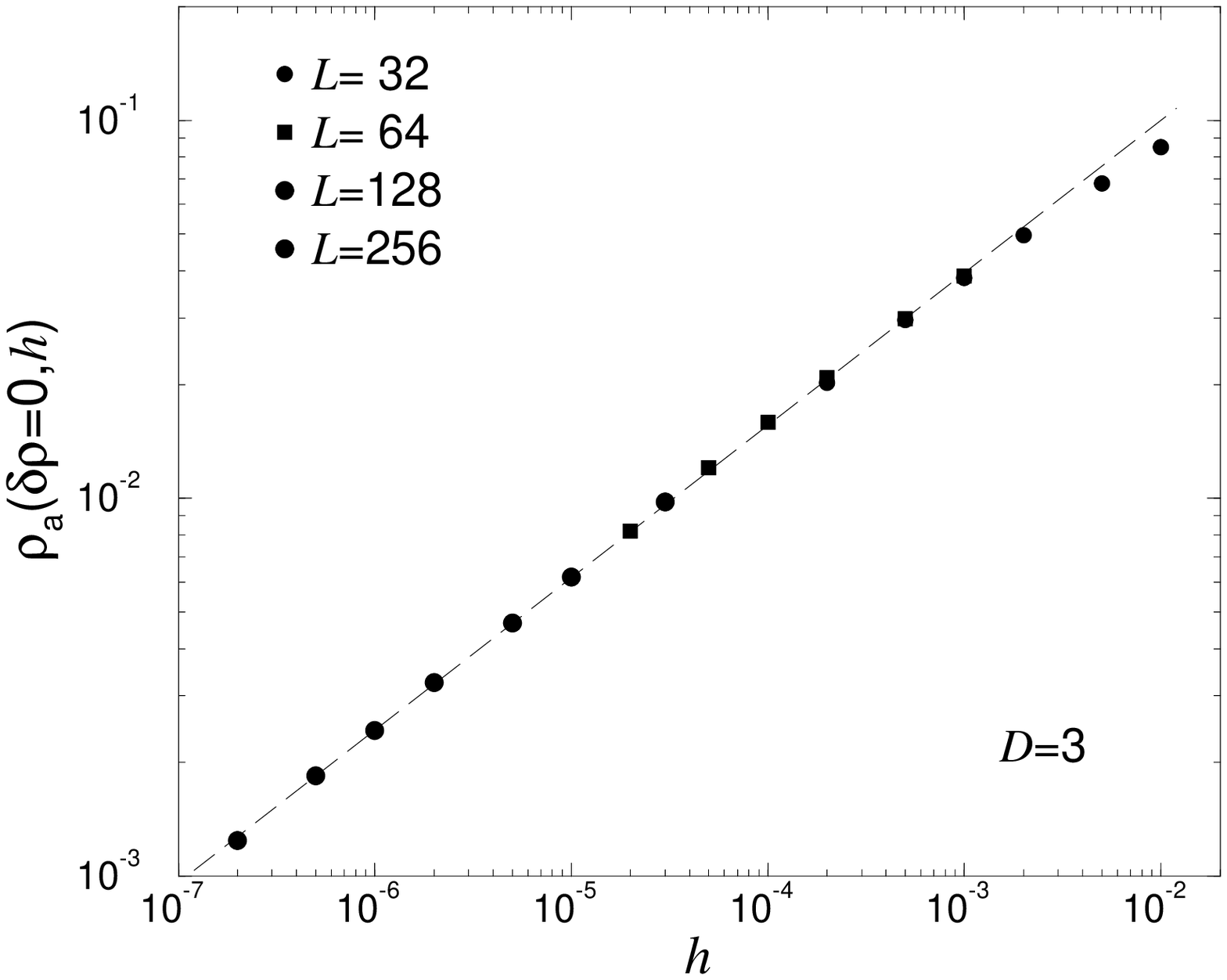}
  \caption{%%
    The order parameter~$\rho_{\text a}$ as a function of the
    field~$h$ at the critical density $\rho=\rho_{\text c}$ for $D=3$.
    The dashed line corresponds to a power-law behavior according
    to Eq.\,(\protect\ref{eq:rho_field}) with an exponent $0.403\pm0.004$.
   }
  \label{fig:rho_a_field_3d_01} 
\end{figure}

\begin{figure}[b]
  \includegraphics[width=8.0cm,angle=0]{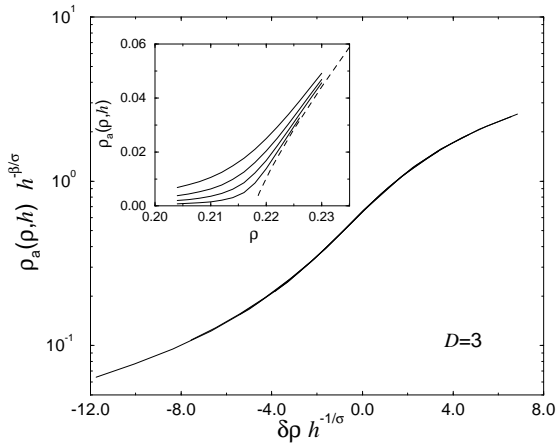}
  \caption{%%
    The scaling plot of the order parameter~$\rho_{\text a}$ for the
    three-dimensional model.
    The data are rescaled according to 
    Eq.\,(\protect\ref{eq:scal_plot_order}) with $\beta=0.837$ and
    $\sigma=2.075$.
    The inset displays the unscaled data, i.e., 
    the order parameter~$\rho_{\text a}$ is plotted as a 
    function of the density $\rho$ for different values 
    of the field~$h$.
    The dashed line corresponds to the zero-field behavior.   }
  \label{fig:rho_a_field_scal_3d_01} 
\end{figure}

Analogous to the two-dimensional case we first determine 
the exponent~$\sigma$ for $D=3$.
The field dependence of the order parameter is plotted
in Fig.\,\ref{fig:rho_a_field_3d_01}.
A regression analysis yields $\beta/\sigma=0.403\pm 0.004$
which leads to the estimation $\sigma=2.075\pm 0.043$
if one uses the value $\beta=0.837\pm 0.015$~\cite{LUEB_19}.

In order to investigate the scaling behavior of the
order parameter we simulated the CLG model for field
values from $h=2\, 10^{-5}$ up to $h=2\, 10^{-4}$.
The corresponding curves as well as the zero-field order
parameter curve is shown in the inset 
of Fig.\,\ref{fig:rho_a_field_scal_3d_01}.
According to Eq.\,(\ref{eq:scal_plot_order}) we plot
the rescaled order parameter in the same figure.
Using $\beta=0.837$ and $\sigma=2.075$ we get again
and excellent data collapse of the different curves.

\begin{figure}[t]
  \includegraphics[width=8.0cm,angle=0]{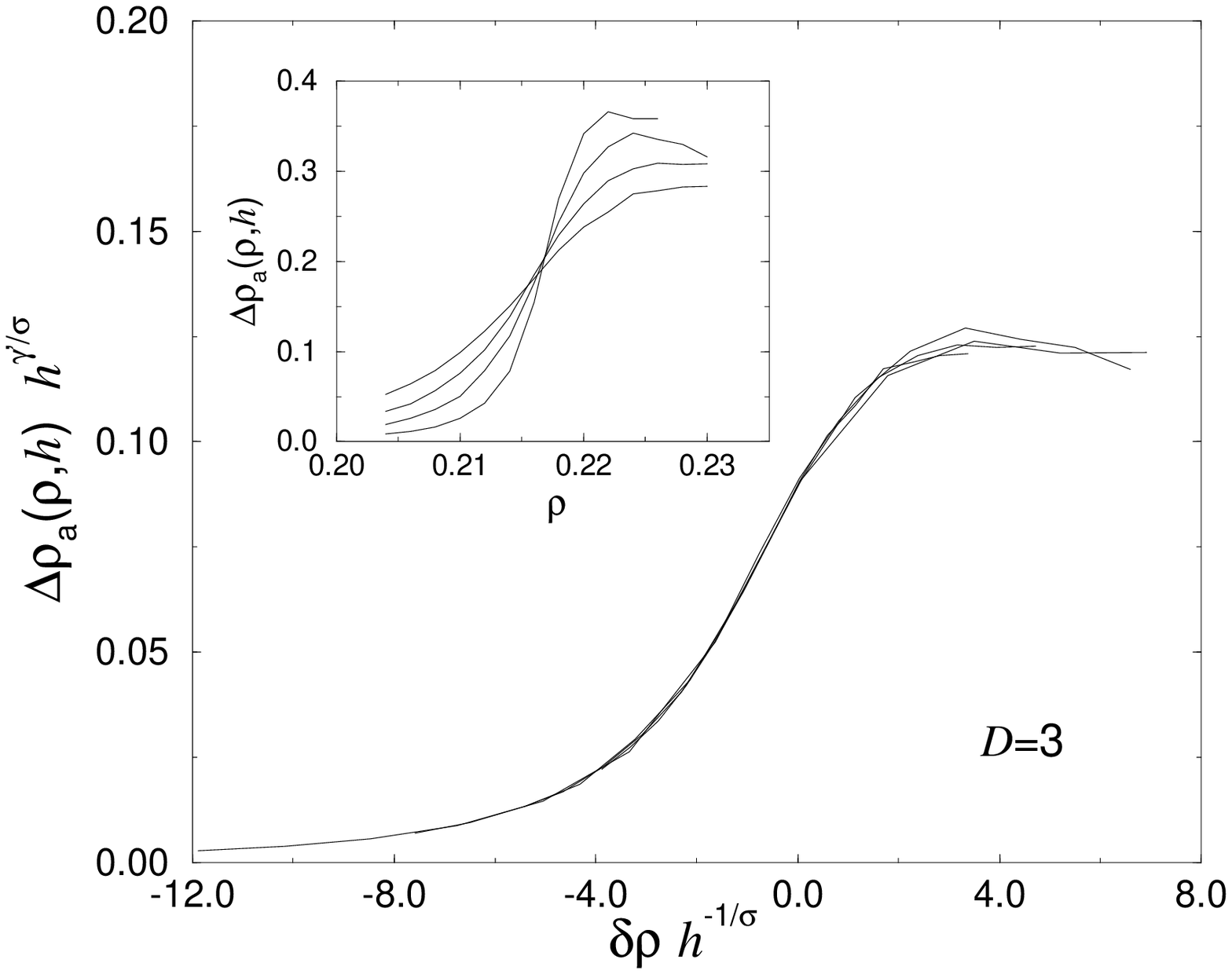}
  \caption{%%
    The scaling plot of the order parameter fluctuations~$\Delta\rho_{\text a}$ 
    for the three-dimensional model.
    The data are rescaled according to 
    Eq.\,(\protect\ref{eq:scal_plot_fluc}) with $\sigma=2.075$ and
    $\gamma^{\prime}/\sigma=0.1$.
    The inset displays the unscaled data, i.e., 
    the fluctuations are plotted as a 
    function of the density $\rho$ for different values 
    of the field~$h$.   
   }
  \label{fig:fluc_a_scal_3d_01} 
\end{figure}

Furthermore we analyse the scaling behavior of the
fluctuations $\Delta \rho_{\text a}(\delta\rho, h)$ for 
the three-dimensional CLG model.
The corresponding data are presented in the inset 
of Fig.\,\ref{fig:fluc_a_scal_3d_01}.
Similar to the two-dimensional case one tries to obtain
a data collapse of these curves by varying the exponent
$\gamma^{\prime}/\sigma$ [Eq.\,(\ref{eq:scal_plot_fluc})].
The best result is obtained for 
$\gamma^{\prime}/\sigma=0.10\pm0.02$ (see Fig.\,\ref{fig:fluc_a_scal_3d_01}) 
which leads to $\gamma^{\prime}=0.208\pm 0.042$.

Previous investigations of the 
fluctuations~$\Delta \rho_{\text a}$ for zero-field
showed that $\Delta \rho_{\text a}$ diverges for
$\delta\rho \to 0$.
But the numerical data could be interpreted either as 
a power-law divergence [Eq.\,(\ref{eq:fluc_crit})]
with an exponent $\gamma^{\prime}=0.18\pm0.06$
or as a logarithmic growth~\cite{LUEB_19}.
Our new results show that the fluctuations of the three-dimensional
CLG model diverges algebraically at the transition point
with the exponent $\gamma^{\prime}=0.208\pm 0.042$.
Additionally, the above results allows to estimate 
the correlation length exponents via the
scaling relation Eq.\,(\ref{eq:scal_rel_gamma_prime})
and one obtains 
$\nu_{\scriptscriptstyle \perp}=0.627\pm0.027$.

\section{At the critical dimension}
\label{sec:dc_d4}

At the upper critical dimension $D=4$ the scaling behavior
of the CLG model is affected by logarithmic 
corrections~\cite{LUEB_19}, i.e., the scaling ansatz 
Eq.\,(\ref{eq:scal_ansatz}) has to be modified.
Motivated by the scaling behavior of the Ising model we 
assume that the order parameter obeys in leading order 
the ansatz (see Appendix)
\begin{eqnarray}
\label{eq:scal_ansatz_dc}
\rho_{\text{a}}(\delta\rho, h) \; = \; & \\ \nonumber 
& \lambda\, |\ln{\lambda}|^l \; {\tilde r}
(\delta \rho \, \lambda^{-1/\beta} |\ln{\lambda}|^b, 
h \, \lambda^{-\sigma/\beta} |\ln{\lambda}|^s), 
\end{eqnarray}
where the exponents $\beta$ and $\sigma$ are given 
by the corresponding mean-field values $\beta=1$ and
$\sigma=2$, respectively.
Thus, for zero field the asymptotic
scaling behavior of the order parameter obeys
\begin{equation}
\rho_{\text{a}}(\delta\rho, h=0) \; \sim \; 
\delta\rho \; | \ln{\delta\rho}|^{\text B}
\label{eq:order_par_dc_zero_field}
\end{equation}
with ${\text B}=b+l$.
This behavior was already observed for the CLG model
with the logarithmic correction exponent 
${\text B} =0.24$~\cite{LUEB_BLABLA}.

\begin{figure}[t]
  \includegraphics[width=8.0cm,angle=0]{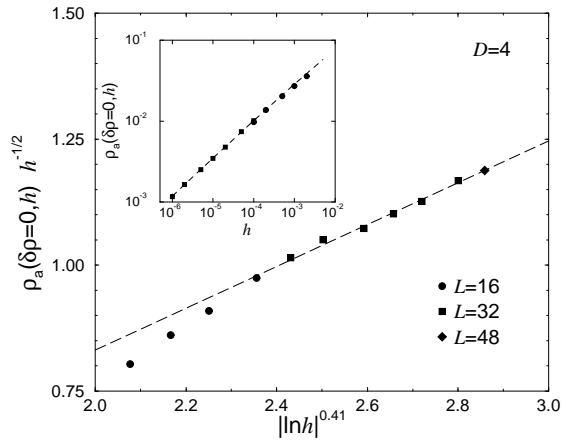}
  \caption{%%
    The order parameter~$\rho_{\text a}$ as a function of the
    field~$h$ at the critical density $\rho=\rho_{\text c}$ for $D=4$.
    According to the scaling ansatz 
    Eq.\,(\protect\ref{eq:order_par_dc_at_crit}) we plot 
    $\rho_{\text a} \, h^{-1/2}$ as a function of 
    $|\ln{h}|^{\Sigma}$.
    The expected asymptotic behavior   
    $\rho_{\text a}\,h^{-1/2}=\text{const}\,|\ln{h}|^\Sigma$
    %[Eq.\,(\protect\ref{eq:order_par_dc_at_crit})]
    is observed for $\Sigma=0.41$ (dashed line).
    The inset displays the original data.
    The dashed line corresponds again to the ansatz 
    Eq.\,(\protect\ref{eq:order_par_dc_at_crit}) with $\Sigma=0.41$.
   }
  \label{fig:rho_a_field_4d_01} 
\end{figure}

According to the above scaling ansatz the asymptotic
field dependence of the order parameter at the critical density
is given by
\begin{equation}
\rho_{\text{a}}(\delta\rho=0, h) \; \sim \; 
h^{1/2} \; | \ln{h}|^{\Sigma}
\label{eq:order_par_dc_at_crit}
\end{equation}
with $\Sigma=s/2+l$.
In our analysis we plot $\rho_{\text a} h^{-1/2}$ as a 
function of $| \ln{h}|^{\Sigma}$ and vary the exponent~$\Sigma$
until one gets asymptotically a straight line.
The best result is obtained for $\Sigma=0.41$ and the corresponding
plot is shown in Fig.\,\ref{fig:rho_a_field_4d_01}.

\begin{figure}[t]
  \includegraphics[width=8.0cm,angle=0]{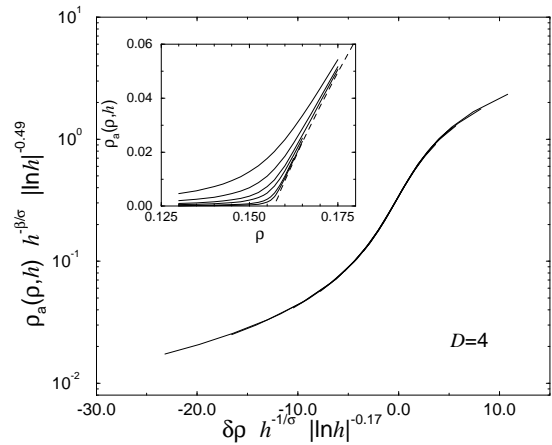}
  \caption{%%
    The scaling plot of the order parameter~$\rho_{\text a}$ for the
    four-dimensional model, i.e., at the upper critical dimension.
    The data are rescaled according to 
    Eqs.\,(\protect\ref{eq:order_par_dc_scal},\protect\ref{eq:scal_arg_lead_ord}) 
    using the mean-field values $\beta=1$ and $\sigma=2$.
    The inset displays the unscaled data, i.e., 
    the order parameter~$\rho_{\text a}$ is plotted as a 
    function of the density $\rho$ for different values 
    of the field~$h$.
    The dashed line corresponds to the zero-field behavior.
   }
  \label{fig:rho_a_field_scal_4d_01} 
\end{figure}

\begin{figure}[b]
  \includegraphics[width=8.0cm,angle=0]{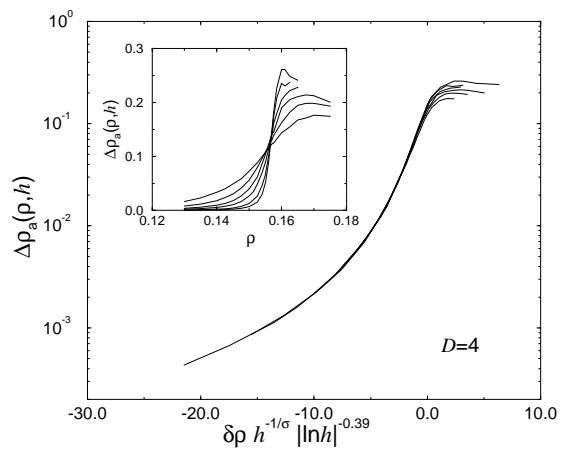}
  \caption{%%
    The scaling plot of the order parameter fluctuations~$\Delta\rho_{\text a}$ 
    for the four-dimensional model.
    The data are rescaled according to 
    Eq.\,(\protect\ref{eq:fluc_scal_ansatz_dc_02}) using the mean-field
    value $\sigma=2$.
    The inset displays the unscaled data, i.e., 
    the fluctuations are plotted as a 
    function of the density $\rho$ for different values 
    of the field~$h$. 
   }
  \label{fig:fluc_a_scal_4d_01} 
\end{figure}

Similar to the lower dimensions we consider the 
scaling behavior of the order parameter as a function
of the control parameter for different external fields.
In the inset of Fig.\,\ref{fig:rho_a_field_scal_4d_01}
we plot $\rho_{\text{a}}$ vs.~$\rho$ for different
fields from $h=10^{-5}$ up to $h=5\,10^{-4}$.
Choosing $h \, \lambda^{-\sigma/\beta} |\ln{\lambda}|^s=1$
the scaling ansatz [Eq.\,(\ref{eq:scal_ansatz_dc})] yields 
in leading order 
\begin{equation}
\rho_{\text{a}}(\delta\rho, h) \; = \; 
h^{1/2} \; | \ln{h}|^{\Sigma} \; 
{\tilde r} (x, 1 ),
\label{eq:order_par_dc_scal}
\end{equation}
where the scaling argument~$x$ 
is given in leading order by
\begin{equation}
x \; = \; 
\delta\rho \; h^{-1/2} \,|\ln{h}|^{b-s/2}
\label{eq:scal_arg_lead_ord}
\end{equation}
Varying the logarithmic correction exponents
one gets a convincing data-collapse, which is shown in 
Fig.\,\ref{fig:rho_a_field_scal_4d_01}, 
for $\Sigma=0.49$ and $b-s/2=-0.17$.
The first value is in good agreement with $\Sigma=0.41$
obtained from the scaling behavior at 
the critical density [Eq.\,(\ref{eq:order_par_dc_at_crit})].
Using the average value $\Sigma=l+s/2=0.45$ and 
$b-s/2=-0.17$ we get the estimation $B=b+l=0.28$ which agrees
with $B=0.24$ obtained from numerical simulations
in zero-field.

Furthermore we consider how the logarithmic corrections affect
the scaling behavior of the fluctuations at the upper
critical dimension.
Similar to the order parameter 
[Eqs.\,(\ref{eq:order_par_dc_scal},\ref{eq:scal_arg_lead_ord})] 
we assume for the leading order of the fluctuations the scaling behavior
\begin{equation}
\Delta \rho_{\text a}(\delta\rho, h) 
\; = \; 
h^{-\gamma^{\prime}/\sigma}\, |\ln{h}|^{\Gamma}\;
{\tilde d}
(\delta \rho \, h^{-1/\sigma}\,|\ln{h}|^{-\eta} , 1)
\label{eq:fluc_scal_ansatz_dc_01}
\end{equation}
with $\sigma=2$.
It is known that the mean-field value of the fluctuation
exponent is $\gamma^{\prime}=0$ which
corresponds to a finite jump of $\Delta\rho_{\text a}$
at the critical density~\cite{LUEB_19,LUEB_20}.
In order to avoid that $\Delta\rho_{\text a}$ diverges
for $h\to 0$ at $\delta\rho=0$ the logarithmic correction
exponent $\Gamma$ has to be set to zero too.
Therefore we try to obtain a data collapse of the fluctuation
data according to the ansatz
\begin{equation}
\Delta \rho_{\text a}(\delta\rho, h) 
\; = \; 
{\tilde d}
(\delta \rho \, h^{-1/2}\,|\ln{h}|^{-\eta} , 1).
\label{eq:fluc_scal_ansatz_dc_02}
\end{equation}
A good data collapse is observed for $\eta=0.39$ and the
corresponding plot is shown in Fig.\,\ref{fig:fluc_a_scal_4d_01}.

\begin{figure}[t]
  \includegraphics[width=8.0cm,angle=0]{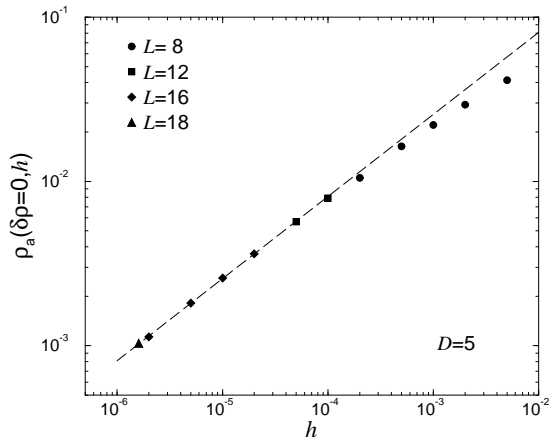}
  \caption{%%
    The order parameter~$\rho_{\text a}$ as a function of the
    field~$h$ at the critical density $\rho=\rho_{\text c}$ for $D=5$.
    The dashed line corresponds to a power-law behavior according
    to Eq.\,(\protect\ref{eq:rho_field}) with $\beta/\sigma=1/2$.
   }
  \label{fig:rho_a_field_5d_01} 
\end{figure}

\section{Above the critical dimension}
\label{sec:d5}

Above the critical dimension, i.e.~$D\ge 5$, the
scaling behavior of the CLG model is expected to
obey again the scaling ansatz Eq.\,(\ref{eq:scal_ansatz})
where the exponents are given by the mean-field values
$\beta=1$, $\sigma=2$, 
and $\gamma^{\prime}=0$~\cite{LUEB_19,LUEB_20},
independently of the particular dimension.

\begin{figure}[t]
  \includegraphics[width=8.0cm,angle=0]{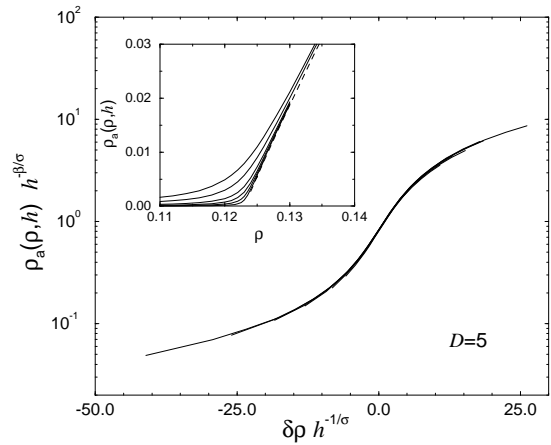}
  \caption{%%
    The scaling plot of the order parameter~$\rho_{\text a}$ for the
    five-dimensional model, i.e., above the upper critical dimension.
    The data are rescaled according to 
    Eq.\,(\protect\ref{eq:scal_plot_order}) using the mean-field values
    $\beta=1$ as well as $\sigma=2$.
    The inset displays the unscaled data, i.e., 
    the order parameter~$\rho_{\text a}$ is plotted as a 
    function of the density $\rho$ for different values 
    of the field~$h$.
    The dashed line corresponds to the zero-field behavior.
   }
  \label{fig:rho_a_field_scal_5d_01} 
\end{figure}

\begin{figure}[b]
  \includegraphics[width=8.0cm,angle=0]{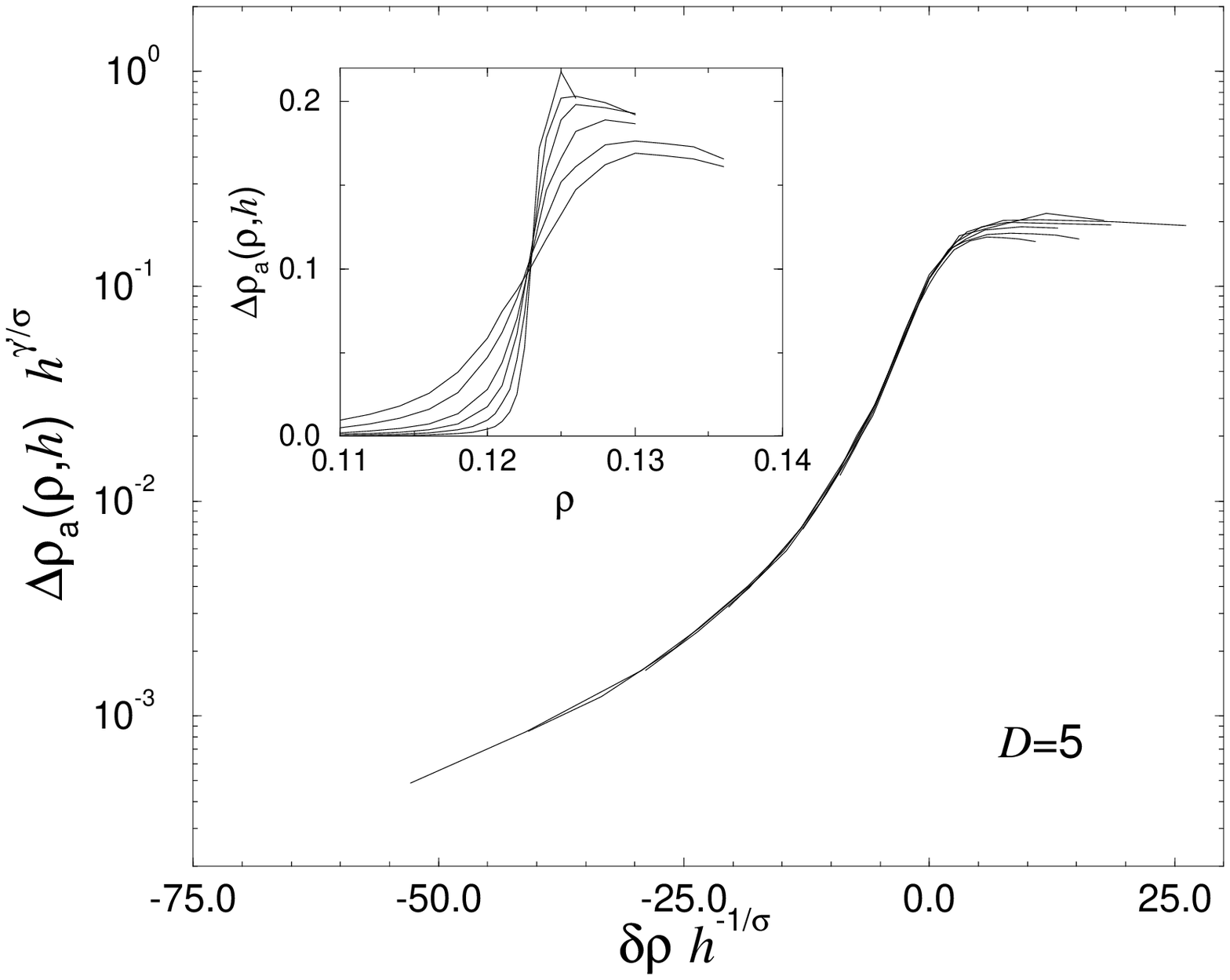}
  \caption{%%
    The scaling plot of the order parameter fluctuations~$\Delta\rho_{\text a}$ 
    for the five-dimensional model.
    The data are rescaled according to 
    Eq.\,(\protect\ref{eq:scal_plot_fluc}) using the mean-field
    values $\sigma=2$ and $\gamma^{\prime}=0$.
    The inset displays the unscaled data, i.e., 
    the fluctuations are plotted as a 
    function of the density $\rho$ for different values 
    of the field~$h$.   
   }
  \label{fig:fluc_a_scal_5d_01} 
\end{figure}

\begin{table*}[t]
\caption{The critical density $\rho_{\text c}$ and 
the critical exponents $\beta$, $\sigma$, $\gamma^{\prime}$ and
$\gamma$ of the CLG model for various dimensions~$D$.
The values of the susceptibility exponent~$\gamma$ are calculated
via Eq.\,(\protect\ref{eq:widom}).
The symbol $^{\ast}$ denotes logarithmic corrections
to the power-law behavior.
The values of $\rho_{\text c}$ and $\beta$ are obtained 
from~\protect\cite{LUEB_19}.}
\label{table:exponents}
\begin{tabular}{llllll}
$\;D$       &  $\;\rho_{\text c}$	& $\;\beta$ &$\;\sigma$         & $\;\gamma^{\prime}$	& $\gamma$ \\  
\colrule \\
$\;2\;$ &  $\;0.34494\;$ & $\;0.637\pm0.009\;$ & $\;2.227\pm0.032\;$ & $\;0.384\pm0.023\;$  & $\;1.590\pm 0.033\;$ \\ 
$\;3$   &  $\;0.21791 $	 & $\;0.837\pm0.015$   & $\;2.075\pm0.043$   & $\;0.208\pm0.042$    & $\;1.238\pm 0.046\;$ \\ 
$\;4$   &  $\;0.15705 $	 & $\;1^{\ast}$        & $\;2^{\ast}$        & $\;0^{\ast}$         & $\;1^{\ast}$         \\ 
$\;5$   &  $\;0.12298 $	 & $\;1$               & $\;2$               & $\;0$                & $\;1$         	  \\      
\end{tabular}
\end{table*}

In Fig.\,\ref{fig:rho_a_field_5d_01} we plot the
field dependence of the order parameter at the
critical density.
For sufficiently small values of the field 
($h\le 10^{-4}$) we
observe the expected scaling behavior 
$\rho_{\text a} \sim h^{1/2}$.
Thus the order parameter  $\rho_{\text a}(\delta\rho, h)$ is
determined for various fields
in this regime (from $h=2\,10^{-6}$ up to
$10^{-4}$).
The curves are shown in the inset of 
Fig.\,\ref{fig:rho_a_field_scal_5d_01}.
Rescaling these curves according to the ansatz 
Eq.\,(\ref{eq:scal_ansatz}) with $\beta=1$ and $\sigma=2$ 
one gets a good data
collapse (see Fig.\,\ref{fig:rho_a_field_scal_5d_01}).

Finally we consider the fluctuations of the 
five-dimensional CLG model.
The inset of Fig.\,\ref{fig:fluc_a_scal_5d_01}
shows how the fluctuations are affected by the 
field~$h$.
With vanishing field the curves become steeper
until one gets a jump~\cite{LUEB_19,LUEB_20} for $h=0$.

Considering the scaling behavior one has to
taken into account that the jump corresponds 
to $\gamma^{\prime}=0$.
Therefore, we plot the fluctuations~$\Delta\rho_{\text a}$ 
as a function of $\delta\rho\, h^{-1/2}$.
The resulting data collapse is shown in 
Fig.\,\ref{fig:fluc_a_scal_5d_01} and confirms the
assumed scaling behavior.

\section{Conclusions}
\label{sec:conc}

We introduced a method which allows to apply an external
field in the CLG model.
The external field obeys the particle conservation
and is conjugated to the
order parameter, i.e., it is realized as a
spontaneous creation of active particles.
We considered the order parameter as well as
its fluctuations of the CLG model
as a function of an external field in various dimensions
($D=2,3,4,5$).
Although the external field drives the system away
from criticality the order parameter obeys 
certain scaling laws for sufficiently small values
of the external field.
These scaling laws are investigated and the corresponding
exponents are determined numerically.
The obtained values of the field exponent~$\sigma$ are 
listed together with other critical indices in 
Table~\ref{table:exponents}.
At the upper critical dimension $D_{\text c}=4$ the usual
scaling behavior has to be modified by additional 
logarithmic corrections.

\acknowledgments
I would like to thank A.~Hucht for
helpful discussions and useful comments on 
the manuscript.

\section{Appendix}
\label{sec:appendix}

Recently, the scaling behavior of the well known Ising
model was investigated at the upper critical dimension
($d_{\text{c}}=4$) and it was argued that the 
singular part of the free energy obeys the finite-size scaling 
ansatz~\cite{AKTEKIN_1}
\begin{equation}
f_L(t,h) \;  = \; L^{-4} \,
{\tilde f}( t L^2 \ln^{1/6}{L}, h L^3 \ln^{1/4}{L})
\label{eq:fss_ising_fenergy}
\end{equation}
where $t$ denotes the reduced temperature 
($t=T/T_{\text c}-1$), $h$ an applied magnetic
field and $L$ denotes the system size.
Thus we assume that the free energy is a generalized 
homogeneous function
\begin{eqnarray}
\label{eq:scal_ising_fenergy}
f(t,h,L)  = & \\ \nonumber 
& \lambda \, 
{\tilde f}( t \lambda^{-1/2} | \ln{\lambda}|^{1/6}, 
h \lambda^{-3/4} |\ln{\lambda}|^{1/4}, L \lambda^{1/4})
%\label{eq:scal_ising_fenergy}
\end{eqnarray}
with $\lambda>0$.
Of course for $\lambda=L^{-4}$ one recovers 
Eq.\,(\ref{eq:fss_ising_fenergy}).
This ansatz can be checked in the following 
way:~the derivative of the free energy with respect 
to the applied field leads to the 
scaling equation of the magnetization
\begin{eqnarray}
\label{eq:scal_ising_mag}
%m(t,h,L)   = & \lambda^{1/4} |\ln{\lambda}|^{1/4} \; \;  \\ \nonumber
%& {\tilde m}( t \lambda^{-1/2} |\ln{\lambda}|^{1/6}, 
%h \lambda^{-3/4} |\ln{\lambda}|^{1/4}, L \lambda^{1/4}) .
m(t,h,L)   =  \lambda^{1/4} |\ln{\lambda}|^{1/4} \; \;  \\ \nonumber
 {\tilde m}( t \lambda^{-1/2} |\ln{\lambda}|^{1/6}, 
h \lambda^{-3/4} |\ln{\lambda}|^{1/4}, L \lambda^{1/4}) .
%\label{eq:scal_ising_mag}
\end{eqnarray}
Choosing $t \lambda^{-1/2} |\ln{\lambda}|^{1/6}=1$
one gets in leading order for the order parameter in the thermodynamic limit
at zero field
\begin{equation}
m(t,h=0)  \sim  t^{1/2} \, |\ln{t}|^{1/3}.
\label{eq:ising_mag_temp}
\end{equation}
The field dependence of the magnetization at the
critical temperature ($t=0$)
\begin{equation}
m(t=0,h)  \sim  h^{1/3} \, | \ln{h}|^{1/3}
\label{eq:ising_mag_field}
\end{equation}
is obtained by setting $h \lambda^{-3/4} |\ln{\lambda}|^{1/4}=1$.
Analogous one gets in leading order for the susceptibility
and specific heat
\begin{equation}
\chi(t,h=0)  \sim  t^{-1} \, |\ln{t}|^{1/3}
\label{eq:ising_suscept}
\end{equation}
and 
\begin{equation}
c(t,h=0)  \sim  |\ln{t}|^{1/3},
\label{eq:ising_spec_heat}
\end{equation}
respectively.
In this way the ansatz Eq.\,(\ref{eq:scal_ising_fenergy})
leads directly to the 
Eqs.\,(\ref{eq:ising_mag_temp},\ref{eq:ising_mag_field},\ref{eq:ising_suscept},\ref{eq:ising_spec_heat}) 
which were already derived in the 1970s by Wegner and Riedel
using renormalization group techniques~\cite{WEGNER_1}.
It is worth to mention that these are exact results  
within the renormalization group theory, i.e., neither the 
values of the mean-field exponents nor the values
of logarithmic correction exponents are obtained 
from approximation schemes like $\epsilon$- or $1/n$-expansions.

In the case of the CLG model we choose a scaling ansatz
for the leading order of the order parameter [Eq.\,(\ref{eq:scal_ansatz_dc})]
with corresponds to Eq.\,(\ref{eq:scal_ising_mag}).


\begin{thebibliography}{10}
\expandafter\ifx\csname natexlab\endcsname\relax\def\natexlab#1{#1}\fi
\expandafter\ifx\csname bibnamefont\endcsname\relax
  \def\bibnamefont#1{#1}\fi
\expandafter\ifx\csname bibfnamefont\endcsname\relax
  \def\bibfnamefont#1{#1}\fi
\expandafter\ifx\csname citenamefont\endcsname\relax
  \def\citenamefont#1{#1}\fi
\expandafter\ifx\csname url\endcsname\relax
  \def\url#1{\texttt{#1}}\fi
\expandafter\ifx\csname urlprefix\endcsname\relax\def\urlprefix{URL }\fi
\providecommand{\bibinfo}[2]{#2}
\providecommand{\eprint}[2][]{\url{#2}}

\bibitem[{\citenamefont{Rossi et~al.}(2000)\citenamefont{Rossi,
  Pastor-Satorras, and Vespignani}}]{ROSSI_1}
\bibinfo{author}{\bibfnamefont{M.}~\bibnamefont{Rossi}},
  \bibinfo{author}{\bibfnamefont{R.}~\bibnamefont{Pastor-Satorras}},
  \bibnamefont{and}
  \bibinfo{author}{\bibfnamefont{A.}~\bibnamefont{Vespignani}},
  \bibinfo{journal}{Phys.\,Rev.\,Lett.} \textbf{\bibinfo{volume}{85}},
  \bibinfo{pages}{1803} (\bibinfo{year}{2000}).

\bibitem[{\citenamefont{{H.\,K.~Janssen}}(1981)}]{JANSSEN_1}
\bibinfo{author}{\bibnamefont{{H.\,K.~Janssen}}}, \bibinfo{journal}{Z.~Phys.~B}
  \textbf{\bibinfo{volume}{42}}, \bibinfo{pages}{151} (\bibinfo{year}{1981}).

\bibitem[{\citenamefont{Grassberger}(1982)}]{GRASSBERGER_2}
\bibinfo{author}{\bibfnamefont{P.}~\bibnamefont{Grassberger}},
  \bibinfo{journal}{Z.~Phys.~B} \textbf{\bibinfo{volume}{47}},
  \bibinfo{pages}{365} (\bibinfo{year}{1982}).

\bibitem[{\citenamefont{L{\protect\"u}beck}(2001)}]{LUEB_19}
\bibinfo{author}{\bibfnamefont{S.}~\bibnamefont{L{\protect\"u}beck}},
  \bibinfo{journal}{Phys.~Rev.~E} \textbf{\bibinfo{volume}{64}},
  \bibinfo{pages}{016123} (\bibinfo{year}{2001}).

\bibitem[{\citenamefont{L{\protect\"u}beck and Hucht}(2001)}]{LUEB_20}
\bibinfo{author}{\bibfnamefont{S.}~\bibnamefont{L{\protect\"u}beck}}
  \bibnamefont{and} \bibinfo{author}{\bibfnamefont{A.}~\bibnamefont{Hucht}},
  \bibinfo{journal}{J.~Phys.~A} \textbf{\bibinfo{volume}{34}},
  \bibinfo{pages}{L577} (\bibinfo{year}{2001}).

\bibitem[{\citenamefont{Hinrichsen}(2000)}]{HINRICHSEN_1}
\bibinfo{author}{\bibfnamefont{H.}~\bibnamefont{Hinrichsen}},
  \bibinfo{journal}{Adv.~Phys.} \textbf{\bibinfo{volume}{49}},
  \bibinfo{pages}{815} (\bibinfo{year}{2000}).

\bibitem[{\citenamefont{Jensen and Dickman}(1993)}]{JENSEN_3}
\bibinfo{author}{\bibfnamefont{I.}~\bibnamefont{Jensen}} \bibnamefont{and}
  \bibinfo{author}{\bibfnamefont{R.}~\bibnamefont{Dickman}},
  \bibinfo{journal}{Phys.~Rev.~E} \textbf{\bibinfo{volume}{48}},
  \bibinfo{pages}{1710} (\bibinfo{year}{1993}).


\bibitem{LUEB_BLABLA}{The logarithmic correction exponent~${\text B=0.39}$ was 
   determined in~\protect\cite{LUEB_19} using the ansatz
   \protect{$\rho_{\text a}\sim (\rho-\rho_{\text c}) \, |\ln{(\rho-\rho_{\text c})}|^{\text B}$}.
   Fitting the same data to the different ansatz {Eq.\,(14)} one gets the 
   value ${\text B=0.24}$.} 

\bibitem[{\citenamefont{Aktekin}(2001)}]{AKTEKIN_1}
\bibinfo{author}{\bibfnamefont{N.}~\bibnamefont{Aktekin}},
  \bibinfo{journal}{J.~Stat.~Phys.} \textbf{\bibinfo{volume}{104}},
  \bibinfo{pages}{1397} (\bibinfo{year}{2001}).

\bibitem[{\citenamefont{{F.\,J.~Wegner} and {E.\,K.~Riedel}}(1973)}]{WEGNER_1}
\bibinfo{author}{\bibnamefont{{F.\,J.~Wegner}}} \bibnamefont{and}
  \bibinfo{author}{\bibnamefont{{E.\,K.~Riedel}}},
  \bibinfo{journal}{Phys.~Rev.~B} \textbf{\bibinfo{volume}{7}},
  \bibinfo{pages}{248} (\bibinfo{year}{1973}).

\end{thebibliography}
\end{document}